\UseRawInputEncoding
\documentclass[twocolumn,showpacs,preprintnumbers,amsmath,amssymb]{revtex4-2}

\usepackage{graphicx}
\usepackage{dcolumn}
\usepackage{bm}
\usepackage{color}
\graphicspath{{./figures_20220621_4_8/}}

\begin{document}

\title{Observing a Phase Transition in a Coherent Ising Machine}

\author{Hiroki Takesue$^{1}$ } 
\email{hiroki.takesue@ntt.com}
 
\author{Yasuhiro Yamada$^{1}$ }
\email{yshr.yamada@ntt.com} 

\author{Kensuke Inaba$^{1}$}
\author{Takuya Ikuta$^{1}$}
\author{Yuya Yonezu$^{1}$}
\author{Takahiro Inagaki$^{1}$}
\author{Toshimori Honjo$^{1}$}

\author{Takushi Kazama$^2$}
\author{Koji Enbutsu$^{2}$}
\author{Takeshi Umeki$^{2}$}
\author{Ryoichi Kasahara$^{2}$}

 \affiliation{%
$^1$NTT Basic Research Laboratories, NTT Corporation, 3-1 Morinosato Wakamiya, 
Atsugi, Kanagawa 243-0198, Japan}
\affiliation{%
$^2$NTT Device Technology Laboratories, NTT Corporation, 3-1 Morinosato Wakamiya, 
Atsugi, Kanagawa 243-0198, Japan}

\date{\today}

\begin{abstract}
The coherent Ising machine (CIM) is known to provide the low-energy states of the Ising model. 
Here, we investigate how well the CIM simulates the thermodynamic properties of a 
two-dimensional square-lattice Ising model. Assuming that the spin sets sampled by the CIM 
can be regarded as a canonical ensemble, we estimate the effective temperature of spins 
represented by degenerate optical parametric oscillator pulses by using maximum likelihood 
estimation. With the obtained temperature, we confirm that the thermodynamic quantities 
obtained with the CIM exhibit phase-transition-like behavior that matches the 
analytical and numerical results better than the mean field approximation does. 
\end{abstract}



\maketitle

Recently, ``Ising machines", which simulate the Ising model by using physical systems that 
substitute for Ising spins, have been drawing attention as a way to circumvent the apparent 
plateau in the progress of digital computers \cite{review}. Among the various optical Ising 
machines demonstrated so far \cite{alireza, ina2, peter, vsdwave, honjo, roma, fabian, baba, 
okawachi,snn} is the coherent Ising machine (CIM) \cite{alireza, ina2, peter, vsdwave, honjo}, 
in which a network of degenerate optical parametric oscillators (DOPO) is used to simulate the 
Ising model. It has been shown that the CIM delivers the low-energy states of the Ising model 
\cite{ina2, peter, vsdwave, honjo}. 

On the other hand, it is important to understand how well the CIM reproduces the characteristics 
of the Ising model. So far, we have simulated a two-dimensional (2D) square-lattice Ising model 
at low temperature on the CIM in order to understand the machine's dynamics in its search for 
the low-energy states \cite{2d}. We have also investigated the behavior of a 2D square lattice in 
an external magnetic field on the CIM and found that its results matched the analytical 
solutions and Monte-Carlo simulations \cite{jiba}. The main motivation of this study is to 
clarify whether a CIM exhibits the thermal phase transition of the ferromagnetic 2D Ising model 
on a square lattice as a statistical system, which was not investigated in our previous studies mentioned above. 
The analytical solution for the 2D square lattice, as conceived by Onsager, 
exhibits a clear phase transition at $(\beta J)_{\rm c} = \frac{\ln (1+\sqrt{2})}{2} \simeq 
0.44$, where $\beta$ and $J$ are respectively the inverse temperature and the spin-spin interaction 
coefficient \cite{onsager}. This analytical solution is based on the assumption that 
the ensemble of Ising spins are canonically distributed in a thermal bath. Therefore, an 
investigation of phase transitions on a CIM may also lead to a better understanding of the 
energy distribution of the ``spins" represented by the DOPOs. 

Another motivation of this work is to see if the performance of the CIM can be simulated using 
mean-field (MF) dynamics, as described in \cite{king, review}. It is known that the MF 
solution to the 2D square lattice Ising model exhibits a phase transition at $(\beta J)_{\rm c, 
MF} = 1/4$, which is clearly different from the exact analytical and numerical solutions. Thus, 
we should be able to learn if there is a difference between the CIM and the ``mean field solver" 
through this study. 

In this Letter, we experimentally demonstrate that the CIM can simulate the phase transition of a 
2D square-lattice Ising model. By varying the amplitude of the injected optical pulses for 
coupling the DOPOs, we could change the effective inverse temperature $\beta$ of the Ising spins represented by the DOPO. As a result, we observed a clear phase transition of the 
magnetization, which matched the theoretical curves of the exact analytical 
solution and a numerical simulation based on the Wang-Landau (WL) method \cite{WL} better than the 
MF approximation. 

\begin{figure*}[thb]
\centerline{\includegraphics[width=.9\linewidth]{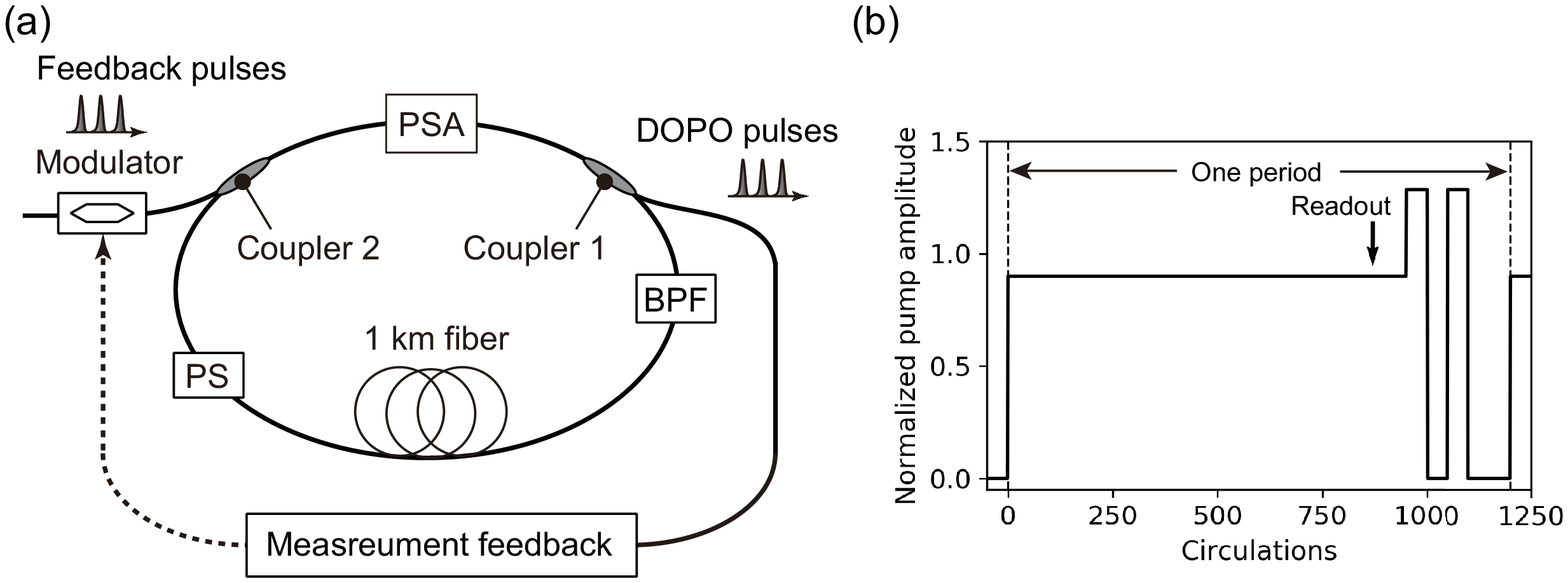}}
\caption{Coherent Ising machine with measurement feedback. (a) Experimental setup. PSA: 
phase-sensitive amplifier. BPF: optical bandpass filter with 0.1-nm bandwidth. PS: phase shifter. 
The measurement-feedback part includes a balanced homodyne detector, an analog-to-digital 
converter, an FPGA, and a digital-to-analog converter. (b) Temporal change in normalized 
pump amplitude for generating the DOPO pulses used for computation as a function of the 
number of circulations in the fiber cavity. The vertical arrow indicates the readout point of the 
DOPO pulses (870th circulation). }
\label{1}
\end{figure*}

Figure \ref{1} (a) shows the experimental setup (for details, see \cite{jiba}). The 
CIM consists of a fiber ring cavity that contains a phase-sensitive amplifier (PSA) based on a 
periodically poled lithium niobate (PPLN) waveguide, two 9:1 fiber couplers, a 0.1-nm width 
optical bandpass filter, a piezo-based phase shifter for cavity locking, and a 1-km optical fiber. 
780-nm, 1-GHz-repetition pump pulses with a $\sim 20$-ps temporal width are launched into 
the PSA. When pumping starts, the PSA generates squeezed vacuum pulses through signal-idler 
degenerate parametric downconversion in the PPLN waveguide. The squeezed vacuum pulses 
circulate in the cavity while undergoing phase-sensitive amplification. Consequently, the 
pulse amplitude saturates after many circulations in the cavity to form 5055 DOPO pulses 
multiplexed in the time domain. Among the 5055 pulses, 512 (referred to as ``signal pulses" 
hereafter) are periodically turned on and off to simulate the given Ising model many times, 
while the remaining pulses oscillate continuously. The DOPO pulses take only 
either 0 or $\pi$ phase relative to the pump phase as a result of the repetitive phase-sensitive 
amplification. By assigning 0 ($\pi$) as spin up (down), the DOPO phase can represent 
an Ising spin state. The ``spin-spin interactions" among the DOPO pulses are implemented with 
the measurement-feedback scheme \cite{ina2,peter}. In this scheme, a portion of the DOPO 
pulse energy is split by Coupler 1 in the ring cavity and one of the quadrature amplitudes is 
measured for all of the DOPO pulses by using a balanced homodyne detector. The measurement 
results are then analog-to-digital converted and input into a fast matrix calculation circuit 
realized with a field programmable gate array (FPGA). Here, the spin-spin interaction matrix for 
a given Ising model of size $N$ is uploaded to the FPGA in advance. The FPGA performs the 
multiplication of the $N \times N$ matrix and an $N$-element vector corresponding to the 
measurement results on the $N$ DOPO pulses, so that a feedback signal for each DOPO pulse 
is obtained in the next round trip. The feedback signal is then used to modulate the amplitude 
and phase of an optical pulse whose wavelength is the same as that of the DOPO pulse in the 
cavity, and the optical pulse is launched into the corresponding DOPO pulse through Coupler 2. 
Here, we denote the amplitude of the injection pulse relative to the DOPO pulses in the cavity 
by $J_{\rm inj}$. We repeat the measurement-feedback procedure for each circulation in the 
cavity. The DOPO pulses circulate 1000 times in the cavity in accordance with the temporal 
schedule of the pump amplitude shown in Fig. \ref{1} (b). The final readout of the DOPO pulse 
amplitudes is undertaken at the 870th circulation. With the current measurement-feedback 
system, we can implement all possible combinations of two-body interactions among 512 signal 
pulses with an 8-bit resolution. 

\begin{figure*}[htb]
\centerline{\includegraphics[width=.95\linewidth]{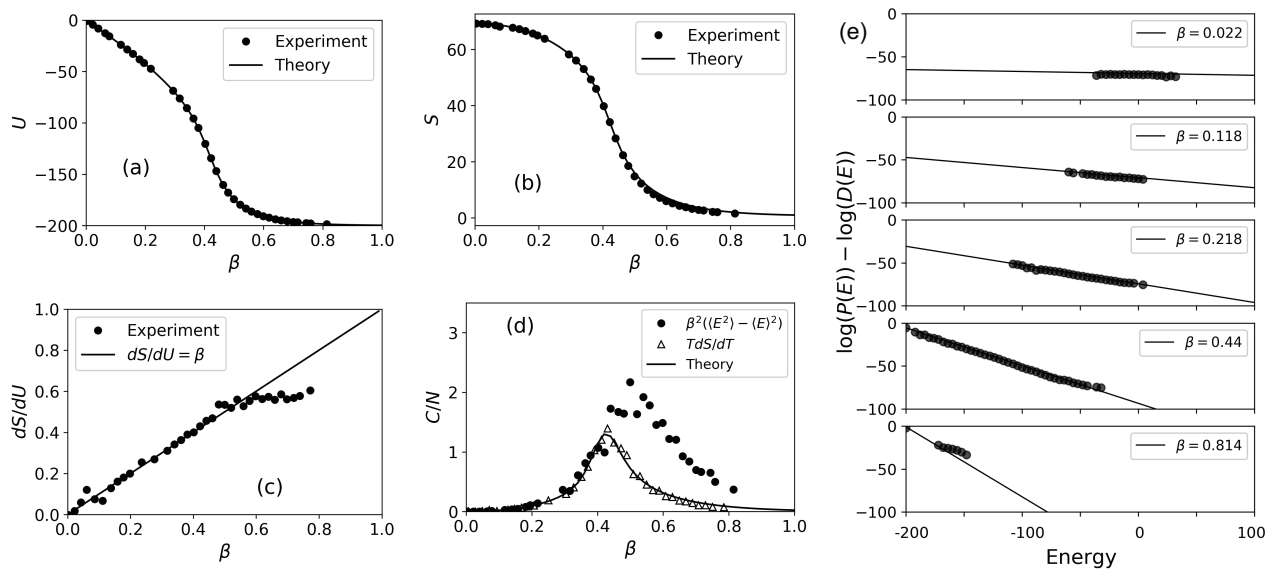}}
\caption{Experimental results. (a) Internal energy $U$ as a function of inverse temperature 
$\beta$. $\beta$ is estimated by fitting the experimentally obtained $U$ to the theoretical 
$U$ curve. (b) Experimentally obtained entropy $S$ as a function of the estimated $\beta$. (c) 
$dS/dU$ as a function of the estimated $\beta$. (d) Specific heat as a function of $\beta$. (e) 
$\ln P(E)/D(E)$ as a function of energy, where $P(E)$ denotes the energy 
distribution at an inverse temperature $\beta$. The lines correspond to theoretical values 
calculated using the estimated $\beta$ and the DOS obtained with the WL method. }
\label{many}
\end{figure*}

We implemented an $N$-spin 2D square lattice with periodic boundary conditions on the CIM. 
The vertical and horizontal links had the same coupling strength. Montroll et al. showed that the 
spontaneous magnetization $M=\sum_i \sigma_i (\sigma_i = \{-1, 1\})$ of the 2D 
square-lattice Ising model has an exact solution at the thermodynamic limit, given by 
\cite{montroll} 
\begin{equation}
M = \left[1-\frac{1}{\sinh^4 2 \beta J} \right]^{1/8}. \label{analytical}
\end{equation}
This equation indicates that the model exhibits a phase transition at $(\beta J)_{\rm c} \simeq 
0.44$. 

\color{black}
In \cite{ina1}, it was empirically shown using one-dimensional (1D) Ising model that the effective temperature of spins represented by DOPO pulses in a CIM depends on the pump amplitude. 
As the pump amplitude is set closer to the oscillation threshold, the defect density decreases, implying a decrease in the effective temperature. 
Similar results were obtained in our numerical simulation based on quantum master equations \cite{yamada}, where not only the pump amplitude but also the injection amplitude altered the effective temperature of the DOPO spins. 
\color{black}
In \color{black} the present \color{black} experiment, the \color{black} optical amplitude of the feedback pulse \color{black} $J_{\rm inj}$ is supposed to be 
proportional to $J$. Therefore, we ran the CIM many times for various $J_{\rm inj}$ values. 
$J_{\rm inj}$ is proportional to the \color{black} electrical \color{black} amplitude of the FPGA output signal, denoted 
by $g$, which can take an integer value in the range between 0 and 127. Consequently, we 
obtained samples of spin sets from which we then obtained the internal energy $U$ (or sample average of the Ising energy) and magnetization values for each $J_{\rm inj}$. 
\color{black} As stated above, the 
increase in $J_{\rm inj}$ also changes the effective inverse temperature $\beta$ of a ``spin" represented by a 
DOPO pulse, \color{black} which means that a change in $J_{\rm inj}$ affects both $\beta$ and $J$. 
Hereafter, we assume that $J=1$ and the effect of a change in $J_{\rm inj}$ is reflected in 
$\beta$. Assuming that the energies are sampled from a canonical distribution, 
the energy distribution at the effective inverse temperature $\beta$, $P(E|\beta)$, and the density of 
states (DOS), $D(E)$, can be related to $\beta$ with the following equation:
\begin{equation} 
P(E|\beta) = D(E) \frac{e^{-\beta E}}{Z}.
\end{equation}
Here, $Z=\sum_E D(E) e^{-\beta E}$ is the partition function, and $D(E)$ of the given graph 
can be calculated using the WL method \cite{WL}. 
Under this assumption, we estimate the inverse temperature $\beta$ by using maximum likelihood 
estimation.
The internal energy at the estimated $\beta$ coincides with the theoretical value $U = \langle E \rangle = \sum_E E P(E|\beta)$.
We plotted the magnetization, entropy $S=-\sum_E P(E) \ln 
\frac{P(E)}{D(E)}$, and specific heat $C$ (defined later) as a function of $\beta$ estimated 
for each sample set, \color{black} where $P(E)$ denotes the experimental energy distribution. \color{black} 
Regarding the magnetization, we used the root mean square (RMS) of 
$M=\frac{1}{N} \sum_i \sigma_i$. 

We implemented a 2D square-lattice Ising model with $N=100$ $ (10 \times 10)$. The 
periodic boundary conditions were implemented by taking advantage of the flexible spin-spin 
interaction enabled by the measurement feedback. We embedded four $N=100$ graphs together 
with a 32-node bipartite graph that was used as a ``monitor graph" to the signal pulses. These graphs were multiplied 
with a 432 $\times$ 432 random permutation and spin flip matrix so that the temporal positions 
and the signs of the spins were randomized to avoid forming ferromagnetic states due to 
experimental imperfections such as reflections of light inside the fiber cavity. 

The CIM performed 256 computations in a single batch of measurements, 
which yielded 1048 spin sets for the 100-node 2D Ising model. In order to extract results 
obtained when the CIM stably operated, we eliminated results when the Ising 
energy of the monitor graph did not reach the ground state. We then used the energies of the 
remaining results in the batch to estimate $\beta$. We set 22 values of the amplitude of the FPGA output signal $g$, ranging from 0 to 125, and obtained ten batches of results for each $g$ value. We 
repeated this procedure four times, which means that we accumulated 40,960 spin sets for each value of $g$. 
On average, $\sim$8\% of the spin sets passed the filtering by the monitor graph. 

Figure \ref{many} (a) shows the internal energy as a function of the estimated $\beta$, which 
fits the theoretical curve as a natural result of the maximum likelihood estimation. The 
experimental entropy as a function of the estimated $\beta$ is shown by the circles in Fig. 
\ref{many} (b). The experimental data clearly fits the theoretical curve, suggesting 
that not only the average energy but also the energy distribution agree with the theory based on 
the canonical distribution assumption. Using the $\beta-U$ and $\beta-S$ relationships, we 
plotted the $S-U$ curve, and obtained the thermodynamic inverse temperature 
defined as $dS/dU$. The relationship between the obtained $dS/dU$ and estimated $\beta$ is 
shown in Fig. \ref{many} (c), which indicates that the estimated temperatures were close to the 
thermodynamic temperature for $\beta$ up to $\sim 0.6$. 

The experimental and theoretical specific heats are shown in Fig. \ref{many} (d). The circles 
show the experimental specific heat derived from the Ising energies and the 
statistical-mechanical relationship $C_{\rm stat} = \beta^2 (\langle E^2 \rangle - \langle E 
\rangle^2)$, and the triangles denote another set of experimental values obtained from the 
estimated entropy (Fig. \ref{many} (b)) and the thermodynamic relationship $C_{\rm thermo} 
= T dS/dT$. The circles form a clear peak structure that is a characteristic of the theoretical 
specific heat as a function of temperature, but at a clearly larger $\beta$. Since $C_{\rm 
stat}$ is proportional to the variance of the energy, the experimental instability 
enhanced at around the transition point may have caused the discrepancy from the theory. On 
the other hand, $C_{\rm thermo}$ (the triangles) fits the theoretical curve very well, 
which suggests that the spin sets produced by the CIM satisfy the thermodynamic relationship. 


\begin{figure}[hb]
\centerline{\includegraphics[width=\linewidth]{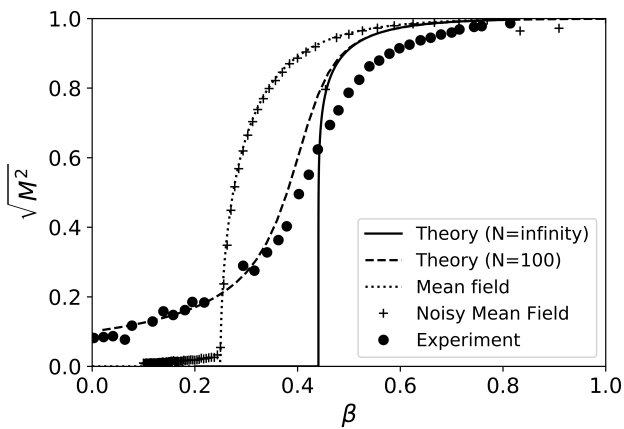}}
\caption{Root-mean-square magnetization as a function of $\beta$. The dots plot the 
experimental spontaneous magnetization as a function of the estimated $\beta$. The solid and 
dashed lines are the theoretical curve for the thermodynamic limit based on Eq. 
(\ref{analytical}) and the numerical calculations based on the WL method for $N=100$. The 
dotted line and $+$ symbols respectively correspond to the MF theory curve and a simulation 
with noisy MF annealing \cite{king}.}
\label{mag}
\end{figure}

Using the estimated $\beta$, we plotted $\ln P(E) - \ln D(E)$ as a function of energy, 
which should be a linear function with slope $\beta$. The results are shown in Fig. \ref{many} 
(e); the experimentally observed distributions fit the theoretical line for the canonical 
distribution at a $\beta$ around the theoretical transition point ($\sim 0.44$) or smaller. 
In contrast, there is a clear discrepancy from the theoretical line at a relatively larger $\beta (\simeq 0.81)$. 
This suggests that the effective temperatures obtained in our experiment are close to the value obtained by statistical mechanics under the canonical distribution assumption, except at temperatures significantly below the transition temperature. Note that this 
tendency resembles the one observed in the $\beta - dS/dU$ curve shown in Fig. \ref{many} 
(c), where the thermodynamic inverse temperatures are smaller than the effective $\beta$ for 
$\beta > \sim0.6$. 
\color{black} 
The validity of the estimated effective temperatures was quantified by the Kullback-Leibler divergence (see Supplemental Material \cite{sm}). 
\color{black} 
We should emphasize here that the present diffusive system satisfies 
the canonical distribution assumption throughout most of the temperature region. The experimentally estimated temperature, thermodynamic temperature, and statistical-mechanics temperature under the canonical distribution assumption are in very good agreement, except at very low temperatures. 



Next, we plotted the RMS magnetization $\sqrt{M^2}$ as a function of the effective $\beta$. 
As is apparent from Fig. \ref{mag}, the experimental result exhibits a clear 
phase-transition-like behavior in the finite-sized system with a critical temperature at $\beta 
\sim 0.42$. In addition, the RMS magnetization curve obtained by the CIM is 
closer than the MF one (dotted line) to the theoretical curve of $M$ in the thermodynamic limit (solid line) and 
the theoretical curve of $\sqrt{M^2}$ for the case of $N=100$ calculated using the DOS 
obtained by the WL method (dashed line). 

Finally, we performed a numerical simulation using an algorithm based on MF dynamics (noisy 
MF annealing) \cite{king}. Here, we 
set the standard deviation of the Gaussian noise $\sigma$ and the amplitude splitting parameter 
$\alpha$ to the same values in \cite{king} (both 0.15). As expected, the results (the $+$ plots in Fig. \ref{mag}) were close to the 
MF solution. These observations indicate that the CIM in its present operational condition well 
simulates the thermodynamic properties of the Ising model, which are presumably hard to 
reproduce with the MF approach. 


It is not yet clear what caused the discrepancy between the estimated temperatures and the 
thermodynamic temperatures at $\beta > \sim0.6$ in Fig.\ref{many} (c).
\color{black} A possible origin of the discrepancy \color{black} is ``mode selection", which was 
observed in our numerical simulation of the CIM \cite{yamada}. In this phenomenon, some 
instances of the Ising model show a tendency that certain spin configurations are more likely to 
appear when the pump amplitude is relatively large. Such ``selected modes" 
\color{black} are visible in the bottom panel of Fig. \ref{many} (e), where the observed energies are discontinuous and do not fit the line that corresponds to the canonical distribution. 
The formation of such modes \color{black} may have prevented the thermodynamic \color{black} inverse \color{black} temperature from \color{black} increasing above \color{black} $\sim 0.6$. Also, the 
mode-selection phenomenon may have caused a deviation from the canonical distribution in the 
low-temperature region, which in turn may have caused the deviation in the experimental 
specific heat $C = \beta^2 (\langle E^2 \rangle - \langle E \rangle^2)$ from theory in Fig. 
\ref{many} (d). 
\color{black}
A similar phenomenon is so-called ``freeze out", which was theoretically discussed in \cite{ryan} and experimentally observed in \cite{2d}. In this phenomenon, when the pump amplitude is sufficiently larger than the oscillation threshold, the DOPO dynamics stop before reaching the ground or low-energy states. 
We suspect that this effect may be the origin of the mode selection. 
Further theoretical and experimental investigation are needed on these issues. 
\color{black}

\color{black}
Recently, several studies on phase transitions simulated on optical Ising machines have been reported \cite{pm1,pm2}.  
 We should point out that the results of these papers matched those of the MF models, whereas our results are in a clear distinction from what is predicted by MF theory. Another fact that differentiates our work from the literature is that we examined the validity of the effective temperature 
by observing not only the spontaneous magnetization but also other physical quantities, namely the entropy and the specific heat, and found that the CIM well simulates a thermodynamic system except at very low temperature.  
MF theory, on the other hand, violates thermodynamic relationships except for a few particular models. 
\color{black} 

In summary, we simulated the phase transition of a 2D square-lattice Ising model on the CIM. 
We confirmed that for relatively small $\beta$, the CIM provides a consistent description of Ising spins in thermal equilibrium from multiple viewpoints, namely thermodynamics, statistical mechanics, and statistical inference. 
With the obtained thermodynamic quantities, we observed phase-transition-like behavior in a finite-sized system with a transition temperature close to the theoretical value. 
These results indicate that the CIM is an optical realization of a thermodynamic spin system where all the spins can be independently accessed. 
We may utilize these characteristics for information processing tasks such as Boltzmann sampling and fast simulation of magnetic systems. 

H. T. and Y. Yamada contributed equally to this work.

\end{document}